\documentclass[twocolumn,showpacs,preprintnumbers,amsmath,amssymb]{revtex4}

\usepackage[unicode]{hyperref}
\usepackage[caption=false]{subfig}
\usepackage{graphics}
\usepackage{epstopdf}
\usepackage{epsfig}
\usepackage{graphicx}
\graphicspath{ {./images/} }
\usepackage{dcolumn}
\usepackage{bm}
\usepackage[caption=false]{subfig}

\begin{document}

\title{On entropy, specific heat, susceptibility and Rushbrooke inequality in percolation 
}%

\author{  M. K. Hassan$^{1}$, D. Alam$^{2}$,  Z. I. Jitu$^{1}$ and M. M. Rahman$^{1}$
}%
\date{\today}%

\affiliation{
$1$ Department of Physics, University of Dhaka, Dhaka 1000, Bangladesh \\
$^2$ Department of Physics, University of Central Florida, Orlando, Florida 32816, USA
}

\begin{abstract}
We investigate percolation, a probabilistic model for continuous phase transition (CPT),
 on square and weighted planar stochastic lattices. In its thermal counterpart, entropy is minimally
low where order parameter (OP) is maximally high and vice versa. Besides, specific heat, 
OP and susceptibility exhibit power-law when approaching the critical point and the corresponding critical exponents 
$\alpha, \beta, \gamma$ respectably obey the Rushbrooke inequality (RI)
$\alpha+2\beta+\gamma\geq 2$. Their analogues in percolation, however, remain elusive. 
We define entropy, specific heat and redefine susceptibility for percolation and show that they behave
exactly in the same way as their thermal counterpart. We also show that RI
holds for both the lattices albeit they belong to different universality classes.
\end{abstract}


\maketitle

The emergence of a well-defined critical value accompanied by a dramatic 
change in the order parameter (OP) and
entropy ($S$) without jump or discontinuity is an indication of second order or continuous phase transition (CPT)
which has acquired a central focus in condensed matter and statistical physics \cite{ref.nishimori}. 
In CPT, the numerical value of OP, which measures the extent of order, is always zero above $T_c$ where 
entropy is significantly high and hence the high-$T$ phase is the disordered phase. On the other hand, 
near $T_c$ the numerical value of $S$, which measures the degree of disorder, 
drops significantly following a sigmoidal shape where OP grows with $\epsilon\sim T-T_c$
 following a power-law ${\rm OP}\sim \epsilon^\beta$ and eventually  $S\rightarrow 0$ 
while ${\rm OP}\rightarrow 1$ revealing that the low-$T$ is the ordered phase.
 The CPT is further characterized by the 
power-law  growth of the specific heat  $C\sim \epsilon^{-\alpha}$ and 
susceptibility $\chi\sim \epsilon^{-\gamma}$ 
near $T_c$ and by their divergences at $T_c$. Remarkably, one finds that a wide range systems belong
to one of a comparatively small number of 
universality classes where each class share the same set of critical exponents. 
Besides this, the values of the critical exponents are bound by some scaling relations. 
One of the most interesting scaling relations is the  Rushbrooke inequality (RI) $\alpha+2\beta+\gamma\geq 2$  
which reduces to equality under static scaling hypothesis \cite{ref.Stanley}.  Many
experiments and exactly solvable models too support equality.

Percolation is one of the simplest paradigmatic model for CPT. 
Besides this, its notion has also been used {\it in extenso} to study spread of forest fire, flow of 
fluid through porous media, spread of biological and computer viruses etc. where the extent of connectivity
has a profound impact \cite{ref.saberi, ref.Newman_virus, ref.Moore_virus}. To define percolation
one has to first choose a lattice or a graph. Then in random bond (site) percolation each bond (site)
 is occupied randomly with probability $p$ independent of the state of 
its neighbors \citep{ref.saberi, ref.Stauffer}. Clearly at $p=0$ in bond percolation, 
each site is a cluster of its own size; while at $p=1$, there is just one
cluster, contiguous sites connected by occupied bonds, of size coinciding with the size of the lattice. 
Interestingly, by tuning $p$ from $p=0$, one finds that clusters
are continuously formed and grown on the average, and eventually arrive at a threshold 
value $p_c$ at which there appears a 
spanning cluster for the first time that spans across the entire system. Such transition from isolated 
finite sized clusters to spanning cluster across $p_c$ is found to be reminiscent of the CPT. This is
why scientists in general and physicists in particular find it so interesting.

In order to make the percolation theory a successful model for CPT, it is necessary that we know how to relate  
its various observable quantities to the corresponding quantities
of the thermal CPT. To this end, Kasteleyn and Fortuin (KF) mapped the $q$-state Potts model 
onto the percolation problem and established some useful connections  \cite{ref.Kasteleyn}.
Thanks to the KF mapping, we now know that the relative size of the spanning cluster
is the order parameter $P$, mean cluster size is the susceptibility,
the relation between $p-p_c$ with linear size $L$ of the lattice is the equivalent counterpart of the relation
between $\epsilon$ and the correlation length $\xi$ etc.
However, we still do not know how to define entropy and specific heat, although they are the key parameters to characterize CPT. 
Besides, we regard the mean cluster size as the susceptibility 
but it exhibits the expected divergence only if the spanning cluster is excluded and even then
the corresponding exponent $\gamma$ is too high to obey RI.
Realizing these drawbacks, some authors already have proposed an alternative 
which although exhibits divergence without the exclusion of the
spanning cluster but the problem of high $\gamma$ still persists
 \cite{ref.fortunato_susceptibility, ref.ziff_susceptibility, ref.bastas_susceptibility, ref.cho}.
Finally, proving whether the Rushbrooke inequality holds 
in percolation or not, remains elusive.

Motivated by the issues outlined above, in this rapid communication,  we 
investigate bond percolation on  square and weighted planar stochastic (WPS) lattices. 
First, we propose a labeled cluster picking probability (CPP) that a site picked
at random belongs to cluster $i$ to measure Shannon entropy $H$.
We show that the resulting entropy has the same expected features as that of its thermal
counterpart. In one phase, where $P=0$, we find that $H$ is significantly high  and
in the other phase where $H\approx 0$ we find that $P$ is significantly high as expected.
Note that $H$ and $P$ both cannot be extremely low or high at the same time since $H$ measures the degree of
disorder and $P$ measures the extent of order.
Second, we define specific heat and find positive critical exponent 
$\alpha$ for both the lattices, which is in sharp contrast to the existing value $\alpha=-2/3$ for square
lattice. We also redefine susceptibility and show that it diverges at the critical point 
without having to exclude the spanning cluster.
Besides, we also obtain its critical exponent $\gamma$ and find that it is significantly smaller 
than the existing known value. The values of $\alpha$ and $\gamma$ re-affirm our earlier findings that 
percolation on square and WPS lattices belong to two distinct universality 
classes although they are embedded in the same spatial dimension \cite{ref.Hassan_Rahman_1}. Finally, 
we find that the elusive RI holds in random percolation regardless of whether it is on 
square or on WPS lattices.

We find it worthwhile first to discuss the construction process of WPS lattice which we proposed
in 2010 \cite{ref.Hassan_Pavel}. It starts with a square of unit
area which we call an initiator. Then, in the first step, we divide the initiator 
randomly into four smaller blocks. In the second step and thereafter, 
only one of the blocks at each step is picked from all the available blocks preferentially 
according to their respective areas and divide that randomly into four smaller blocks. 
The details of the algorithm and image of the lattice can be found in \cite{ref.Hassan_Pavel, ref.Hassan_Dayeen}. 
Percolation on such a lattice has already shown unique results \cite{ref.Hassan_Rahman_1}.
For instance, it is well-known that random percolation on all 
lattice, regardless of the type of percolation and the structural difference of the lattice, 
 share the same set of critical exponents {\it vis-a-vis} belong to the same universality class provided they share 
the same dimension. 
However, we have recently shown that percolation on WPS lattice  does not belong to the
universality class of all the known two dimensional lattices \citep{ref.Hassan_Rahman_1}. 

To study percolation, we use the Newman-Ziff (NZ) algorithm which,
 apart from being the most efficient one, has also many other advantages \cite{ref.Ziff}. For instance,  
it helps calculating various observable quantities over the entire range of $p$ in every 
realization instead of measuring them for a fixed probability $p$ in each realization. 
According to the NZ algorithm, all the labeled bonds $i=1,2,3,..., M$ are first randomized
and then arranged in the order in which they will be occupied. Using the periodic boundary condition we get
$M=2L^2$ for square lattice and $M\sim  8t$ for WPS 
lattice where $t$ is the time step. One advantage of using NZ algorithm is that we 
can create percolation states consisting of $n+1$ 
occupied bonds simply by occupying one more bond to its immediate past state consisting of $n$ occupied 
bonds. Initially, there are $L^2$ and $3t+1$ clusters of size one in the square and WPS lattices respectively.
Occupying the first bond means forming a cluster of size two. Each time thereafter, either 
the size of an existing cluster grows due to occupation of inter-cluster bond or remains 
the same due to occupation of intra-cluster bond. We calculate an observable, say $X_n$, as 
a function of $n$ and use it in the convolution relation
\begin{equation}
\label{eq:convolution}
X(p)=\sum_{n=1}^M \left( \begin{array}{c}
M \\ n \end{array}\right ) p^n(1-p)^{M-n} X_n,
\end{equation}
to obtain $X$ as a function of $p$ that helps obtaining a smooth curve for $X(p)$.

Phase transitions always entails a change in entropy in the system  
regardless of whether the transition is first or second order in character.
Thus, a model for CPT is not complete without a proper definition of entropy. 
We find that the Shannon entropy is the most appropriate 
one for percolation as it is too probabilistic in nature. In general, it is defined as
\begin{equation}
\label{eq:shannon}
H=-K\sum_i^m \mu_i\log \mu_i,
\end{equation} 
where we set the constant $K=1$ since it merely amounts to a choice of a unit of measure of entropy \cite{ref.shannon}.
In information theory, it is a common practice to choose $K=1/\log 2$.
  Although there is no explicit restriction per se on the choice of $\mu_i$ for Shannon entropy but when we 
use it to describe entropy for phase transition
there ought to have some implicit restrictions.  We all know from thermodynamics and statistical mechanics 
that $G$ vs $T$ should be a concave curve with
negative slope to ensure that entropy $S\geq 0$. On the other hand, the slope  of the $S$ vs $T$ plot must 
have a sigmoid shape with positive slope since $C\geq 0$ 
and according to second law of thermodynamics $\Delta S\geq 0$ \cite{ref.Stanley}. 
The entropy for percolation too must have the same generic features.

\begin{figure}
\centering
\subfloat[]
{
\includegraphics[height=4.0 cm, width=2.4 cm, clip=true,angle=-90]
{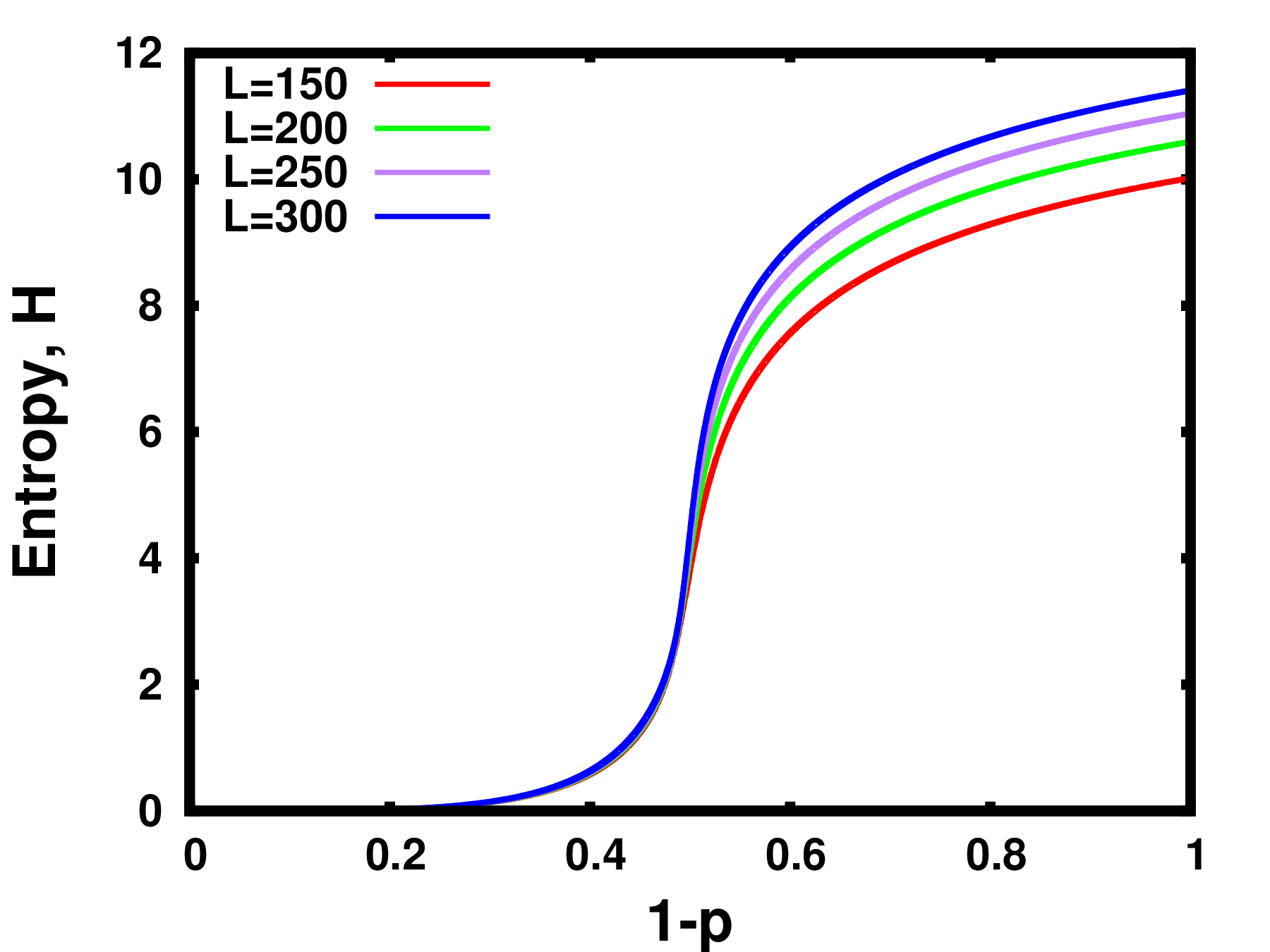}
\label{fig:1a}
}
\subfloat[]
{
\includegraphics[height=4.0 cm, width=2.4 cm, clip=true, angle=-90]
{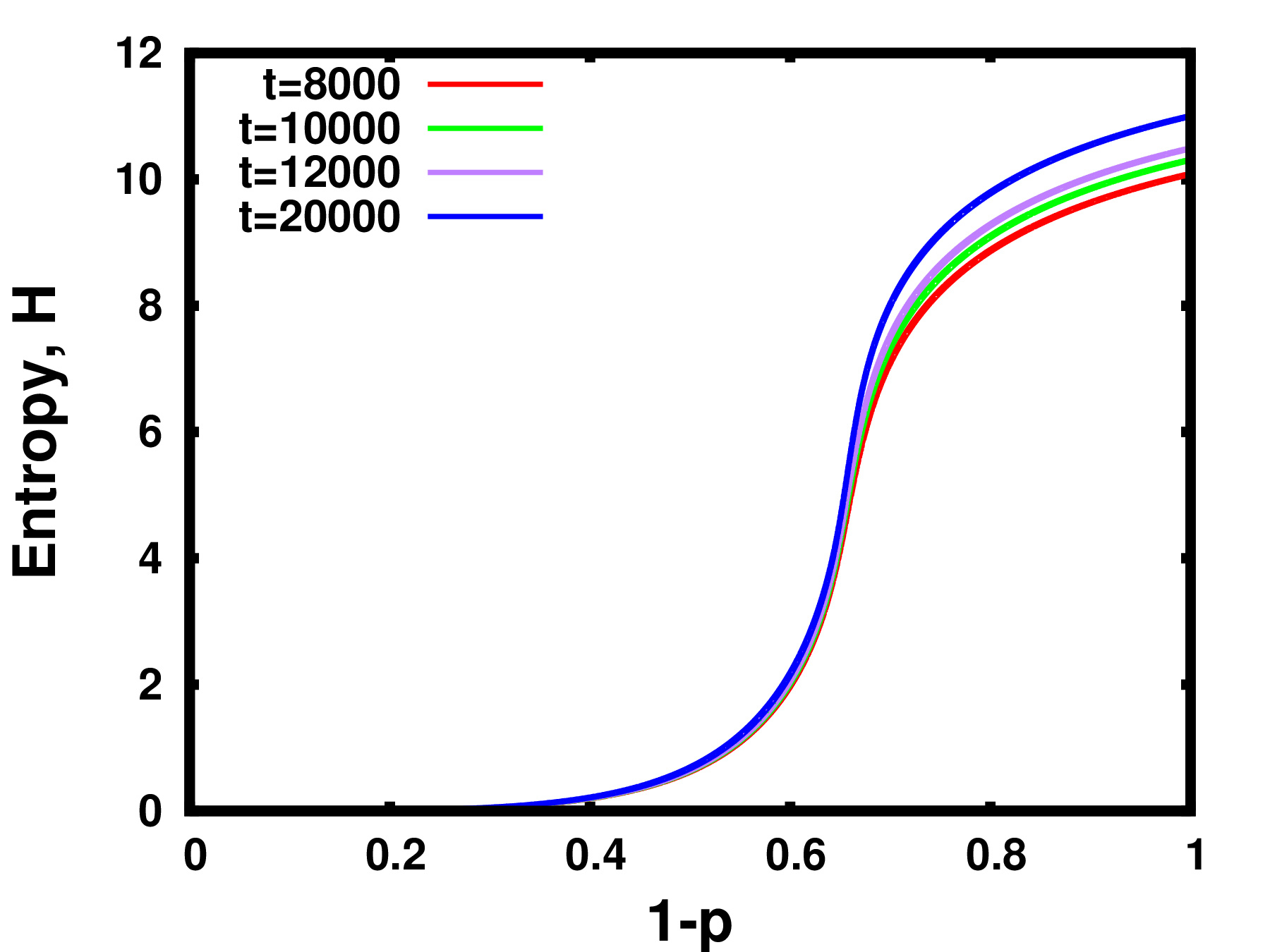}
\label{fig:1b}
}

\caption{Change in Shannon entropy $H$ with $1-p$ for (a) square and (b) WPS lattices. The sharp rise in $H$
occurs near $q_c=0.5$ and $q_c=0.6543$ for (a) and (b) respectively. 
} 
\label{fig:1ab}
\end{figure}

The key question in percolation is not whether we can measure Shannon entropy or not. Rather, faced with
a number of different normalized probabilities, the question is, can we use them?  Earlier, some authors used $w_s$, the probability that  
a site picked at random belongs to a cluster exactly of size $s$, to 
measure entropy using Eq. (\ref{eq:shannon}) and they both found a bell-shaped curve with peaks at around $t_c$ \cite{ref.tsang, ref.vieira}. 
It implies that the system is in the most ordered state at $p=0$ since entropy is minimally low there and at the same time it is in the most disordered
state since the order parameter $P$ is also zero there and hence we see a contradiction. The problem lies in using $w_s$ is that the sum in 
Eq. (\ref{eq:shannon}) is not over cluster size $s$ rather it is over cluster label $i$ so that it measures the amount of information conveyed by each cluster
not by a class or group of cluster of size $s$. Thus, we have to choose a probability that 
contains information about individual clusters.
To find the appropriate probability for $\mu_i$ of Eq. (\ref{eq:shannon}), we assume that for a given
$p$ there are $m$ distinct, disjoint, and indivisible labeled clusters $i=1,2,...,m$ of size $s_1,s_2,....,s_m$ 
respectively. We then propose the labeled picking probability (CPP) $\mu_i$, that a
 site picked at random belongs to cluster $i$. The most generic choice for CPP would be $\mu_i\propto s_i$ so that the probability 
 $\mu_i(p)=s_i/\sum_{j=1}^m s_j$ where $\sum_{j=1}^m s_j=N$ is
the normalization factor.

Substituting $\mu_i(p)=s_i/N$ in Eq. (\ref{eq:shannon})  we first find microcanonical ensemble average of 
entropy $H_n$ as a function of $n$ from $10,000$ independent
realizations. Then we use it in Eq. (\ref{eq:convolution}) to get the canonical ensemble average 
of $H(p)$ as a function of $p$. 
In Figs. \ref{fig:1a} and \ref{fig:1b} we plot $H(p)$ versus $q=1-p$ and find that the curve has the 
desired sigmoidal shape. We see that
the entropy is maximum $H=\log(N)$ at $q=1$ where $\mu_i=1/N$ $ \forall \ i$ as there are $N$ clusters 
of equal size  (one). This is exactly like the state of an isolated 
ideal gas where all microstates are equally likely and hence $q=1$ corresponds to
the most disordered  state and it is consistent with the fact that $P=0$ there. 
As we lower the $q$ value from $q=1$ we 
see $H$ decreases slowly, however, as $q$ approaches $q_c$ we observe a sudden drops in entropy. 
This is because in the vicinity of $p_c$ 
coagulation of two moderately large cluster happens so frequently that we already see a sign of the emergence of 
would spanning cluster that
it causes a sharp rise of CPP which eventually becomes the spanning cluster.  
Already at $q_c$ the incipient spanning cluster becomes so 
prevailing that it embodies almost all the sites. Lowering $q$ further to below $q_c$, we see that
the extent of uncertainty diminishes sharply. Exactly at $q=0$,  we have $\mu_1=1$ and hence entropy $H=0$.
This situation is like perfect crystal as the extent of uncertainty, which is synonymous to degree of disorder, 
completely ceases to exist. The term percolation thus refers to the transition across $q_c$ from ordered phase at low-$q$  to 
disordered phase at high-$q$ exactly like ferromagnetic transition.

\begin{figure}
\centering

\subfloat[]
{
\includegraphics[height=4.0 cm, width=2.5 cm, clip=true,angle=-90]
{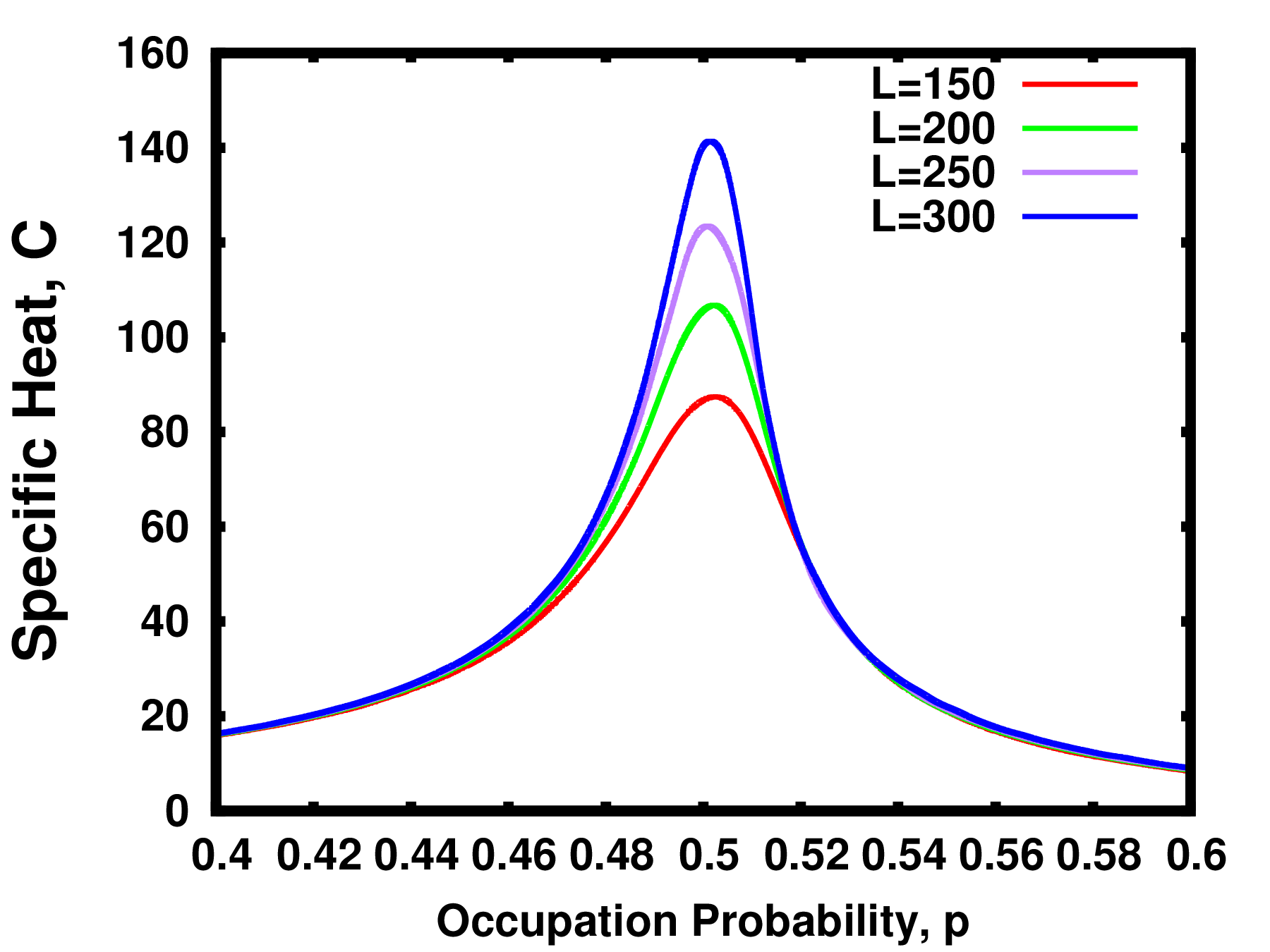}
\label{fig:2a}
}
\subfloat[]
{
\includegraphics[height=4.0 cm, width=2.5 cm, clip=true, angle=-90]
{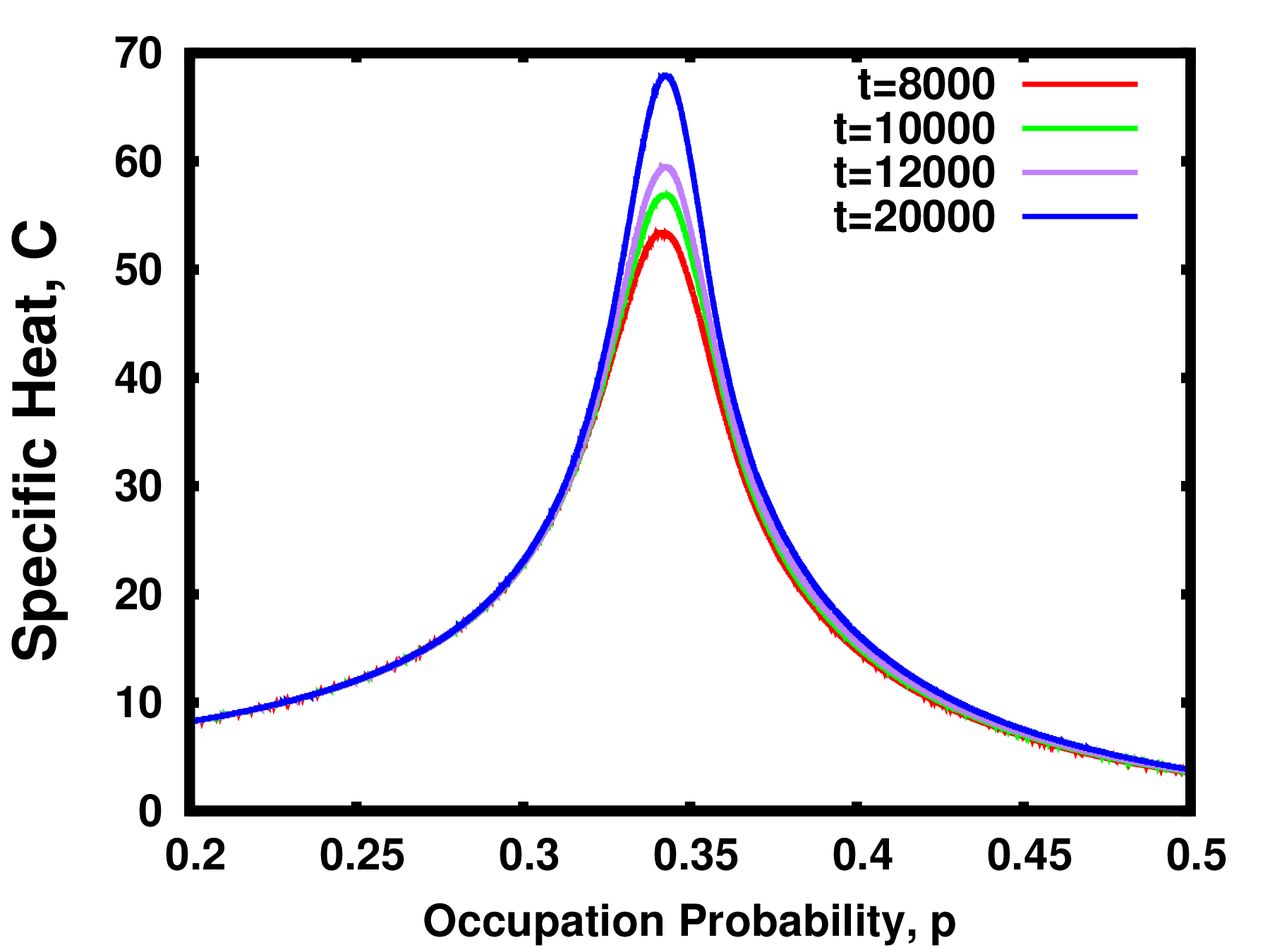}
\label{fig:2b}
}

\subfloat[]
{
\includegraphics[height=4.0 cm, width=2.5 cm, clip=true, angle=-90]
{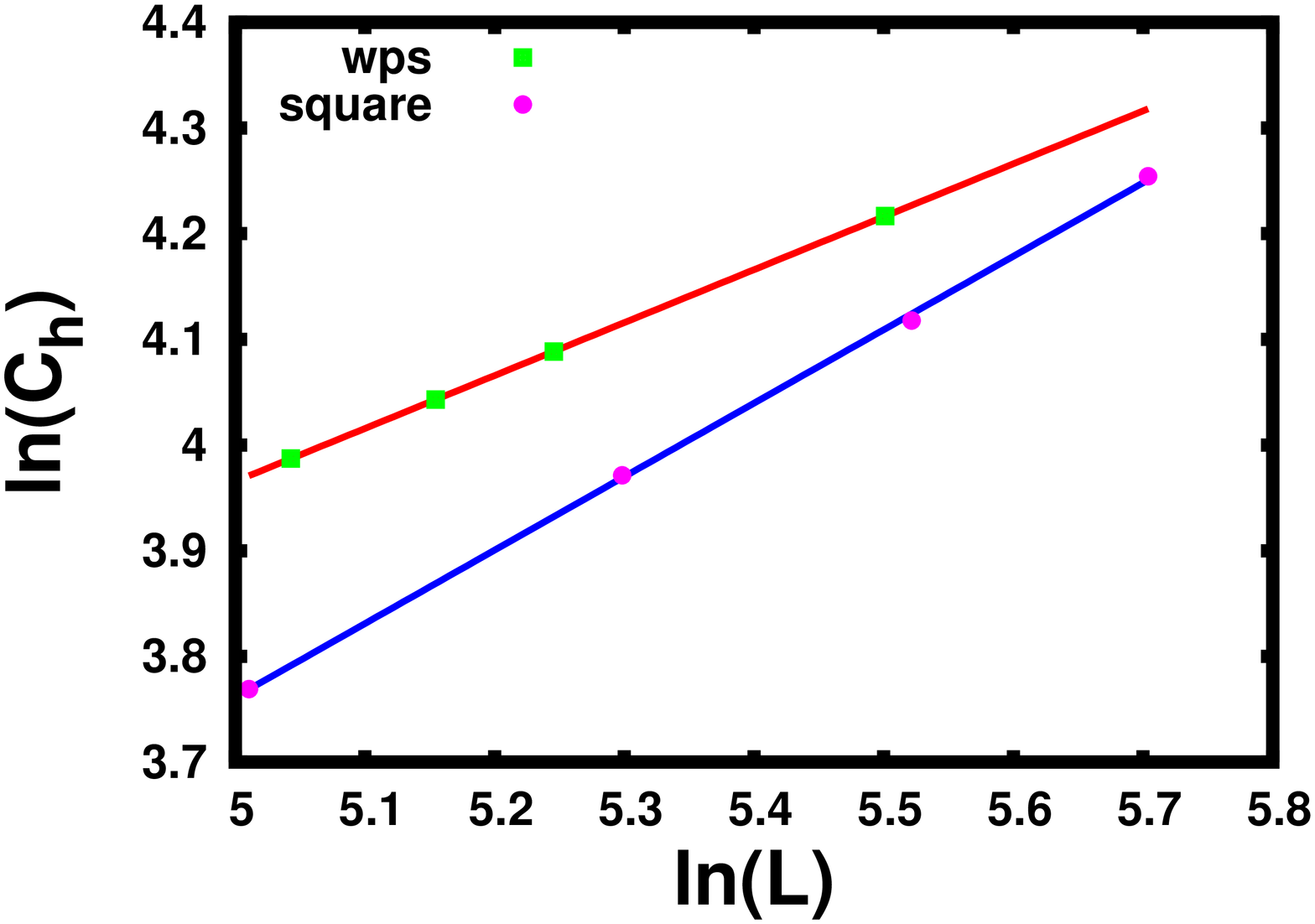}
\label{fig:2c}
}
\subfloat[]
{
\includegraphics[height=4.0 cm, width=2.5 cm, clip=true, angle=-90]
{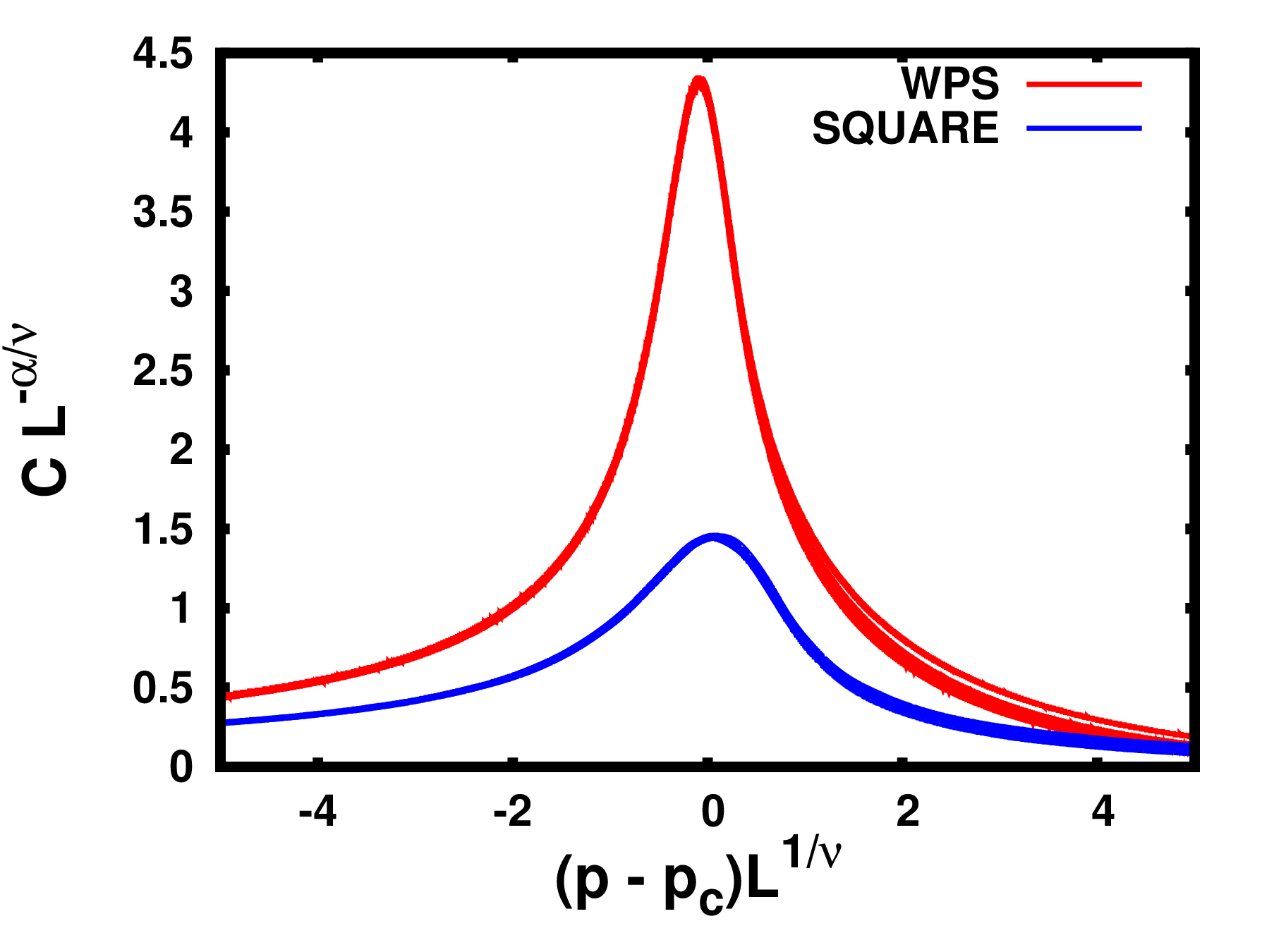}
\label{fig:2d}
}

\caption{
Specific heat $C(p)$ vs $p$ for (a) square and (b) WPS lattices. 
(c) Slope of the plots of $\log(C_{{\rm h}})$ versus $\log(L)$ gives $\alpha/\nu=0.68(1)$ and $\alpha/\nu=0.5007(3)$ for square and WPS lattices respectively.
(d) Plots of $C(p)L^{-\alpha/\nu}$ vs $(p-p_c)L^{1/\nu}$ of the same data as in (a) and (b) shows excellent data-collapse.
} 
\label{fig:3abcd}
\end{figure}

To find the specific heat $C(p)$ for percolation we just need to use its thermodynamic definition $C=T{{dS}\over{dT}}$ 
and replace $S$ by $H$ and $T$ by $1-p$ to obtain
\begin{equation}
C(p)=(1-p){{dH}\over{d(1-p)}}.
\end{equation}
In Figs. \ref{fig:2a} and \ref{fig:2b} we plot it for both the lattices as a 
function of $p$ for different system sizes. Now following the finite-size scaling (FSS) hypothesis we can write
\begin{equation}
\label{eq:specific_heat_fss}
C(p,L)\sim L^{\alpha/\nu}\phi_C((p-p_c)L^{1/\nu}),
\end{equation} 
where $\phi_C$ is the scaling function for specific heat.
To find an estimate for the exponent $\alpha/\nu$, we measure the 
height of the peak $C_{{\rm h}}$ at $p_c$ as a function of $L$.
Plotting $\log(C_{{\rm h}})$ versus $\log(L)$ we get straight lines, see Fig. (\ref{fig:2c}),  
with slopes $\alpha/\nu=0.68(1)$ for square lattice and $\alpha/\nu=0.5007(3)$ for WPS lattice. 
Note that  $CL^{-\alpha/\nu}$ and $(p-p_c)L^{1/\nu}$ are dimensionless quantities and hence 
if we now plot $CL^{-\alpha/\nu}$ as a function of $(p-p_c)L^{1/\nu}$ then the distinct plots of $C$ vs $p$ 
should collapse onto a single universal curve $\phi_c$. Indeed, Fig. (\ref{fig:2d})
shows that all the distinct plots 
 of Figs. (\ref{fig:2a}) and (\ref{fig:2b}) are collapse superbly 
into their own universal curve revealing that if we had data for infinite 
Using the relation  $L\sim (p-p_c)^{-\nu}$ in $C(p)\sim L^{\alpha/\nu}$ we 
find that the specific heat diverges
\begin{equation}
C(p)\sim (p-p_c)^{-\alpha},
\end{equation} 
with $\alpha=0.906(13)$ for square lattice and $\alpha=0.816(4)$ for WPS lattice.
Our value of $\alpha$ for square lattice is in sharp contrast to the existing value $\alpha=-2/3$ \cite{ref.domb}.
Note that the negative value of $\alpha$ means specific heat does not diverge at the critical
point rather it behaves more like order parameter.

\begin{figure}
\centering
\subfloat[]
{
\includegraphics[height=4.0 cm, width=2.4 cm, clip=true,angle=-90]
{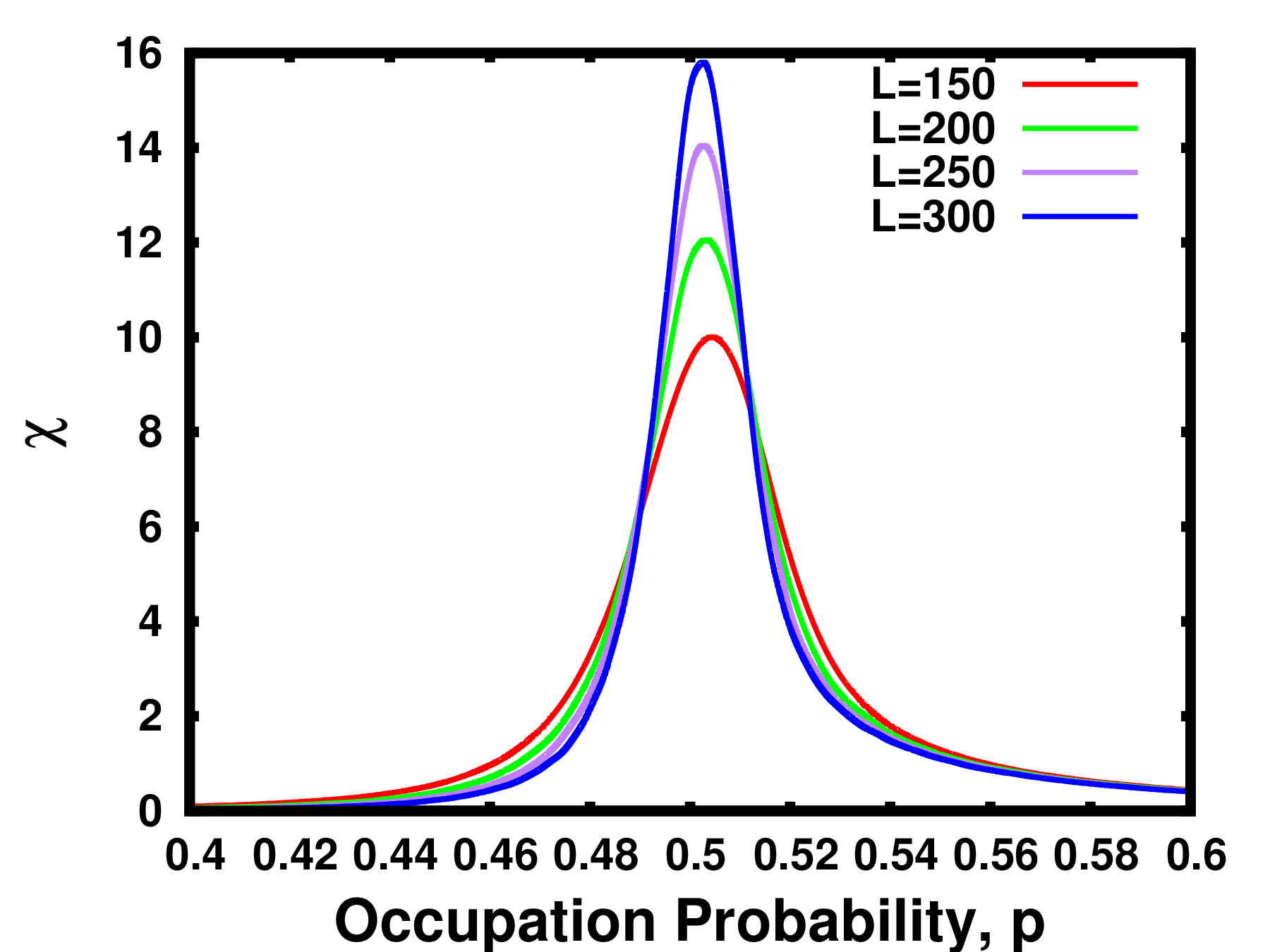}
\label{fig:3a}
}
\subfloat[]
{
\includegraphics[height=4.0 cm, width=2.4 cm, clip=true,angle=-90]
{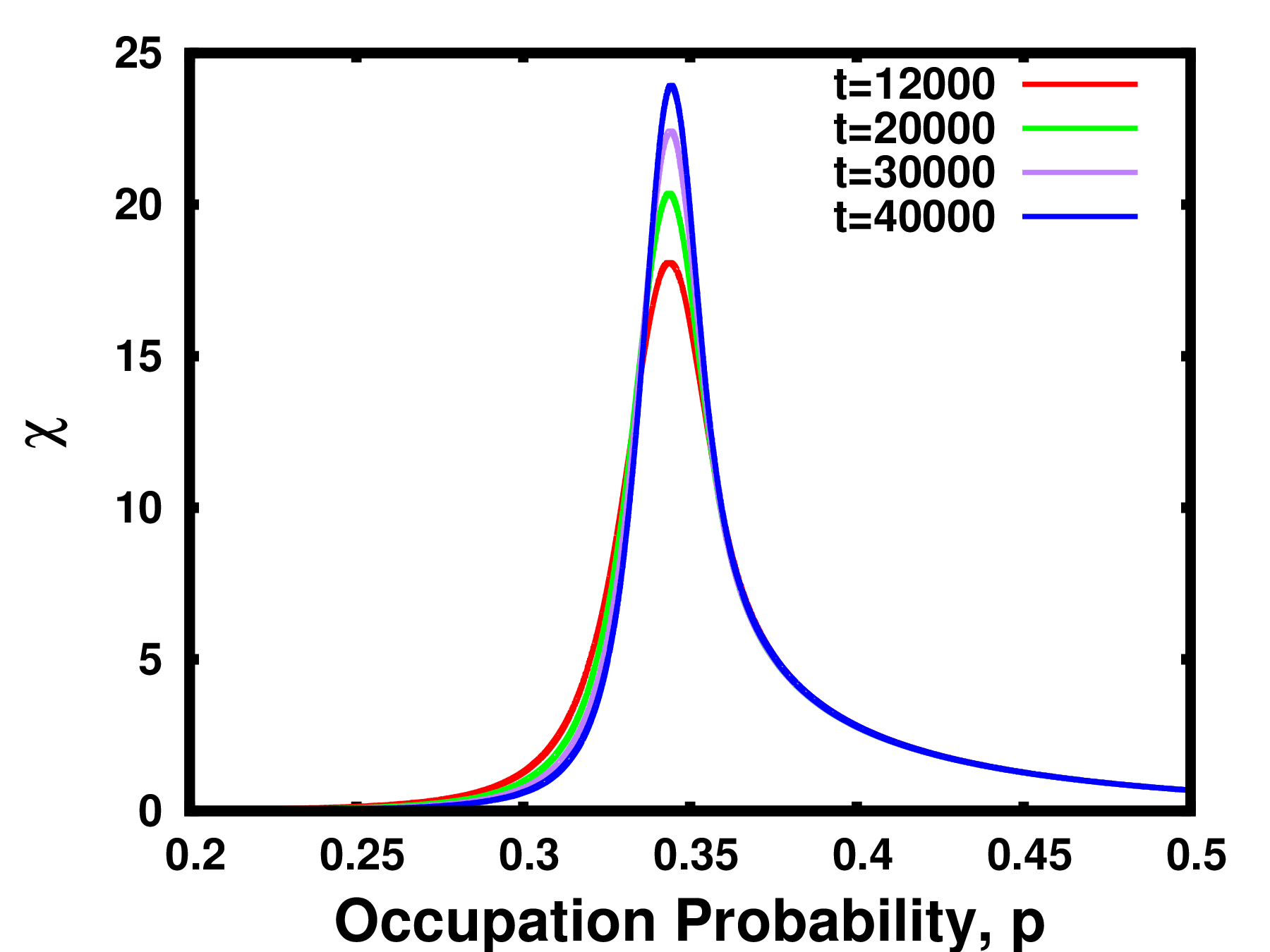}
\label{fig:3b}
}

\subfloat[]
{
\includegraphics[height=4.0 cm, width=2.5 cm, clip=true, angle=-90]
{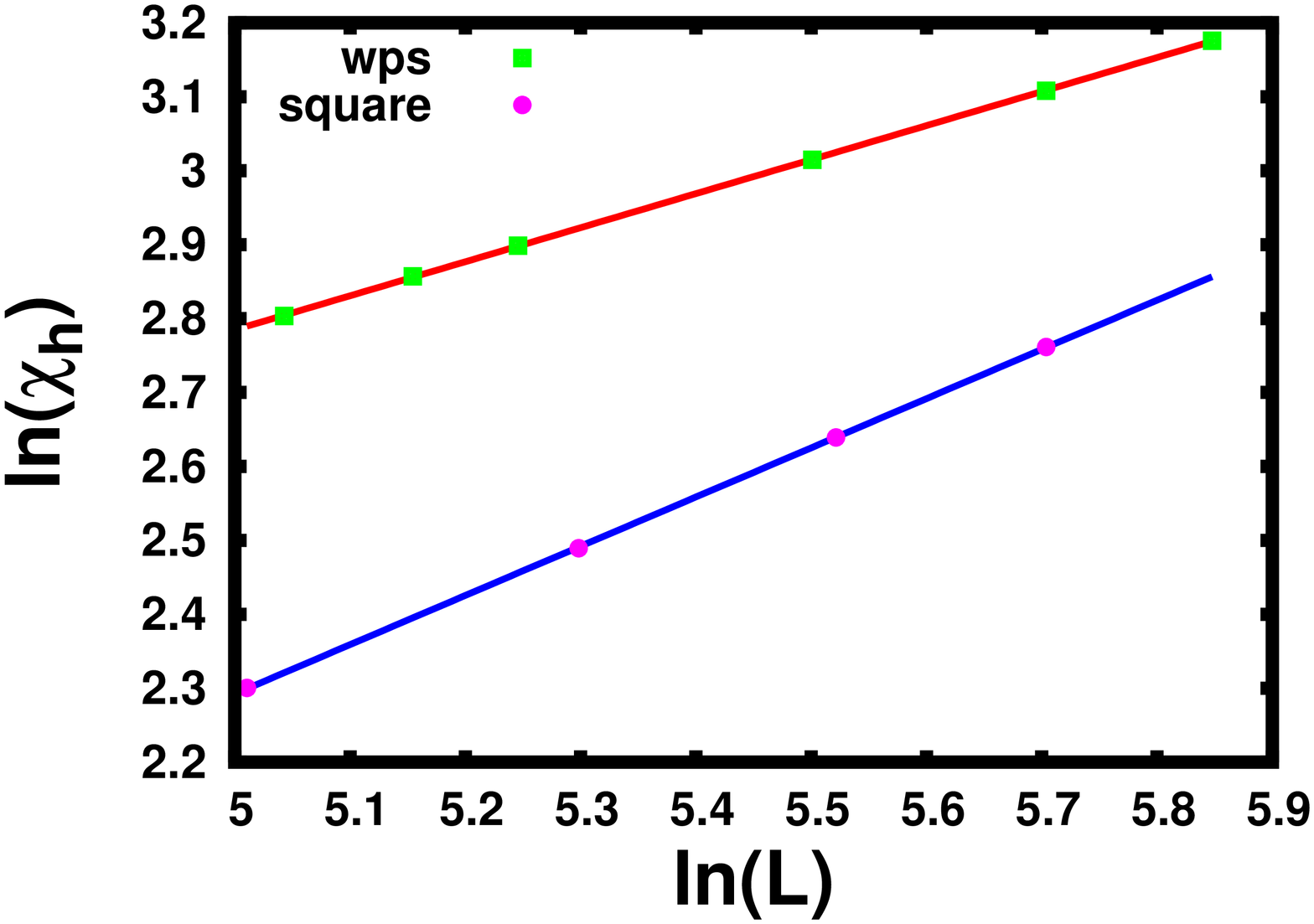}
\label{fig:3c}
}
\subfloat[]
{
\includegraphics[height=4.0 cm, width=2.4 cm, clip=true, angle=-90]
{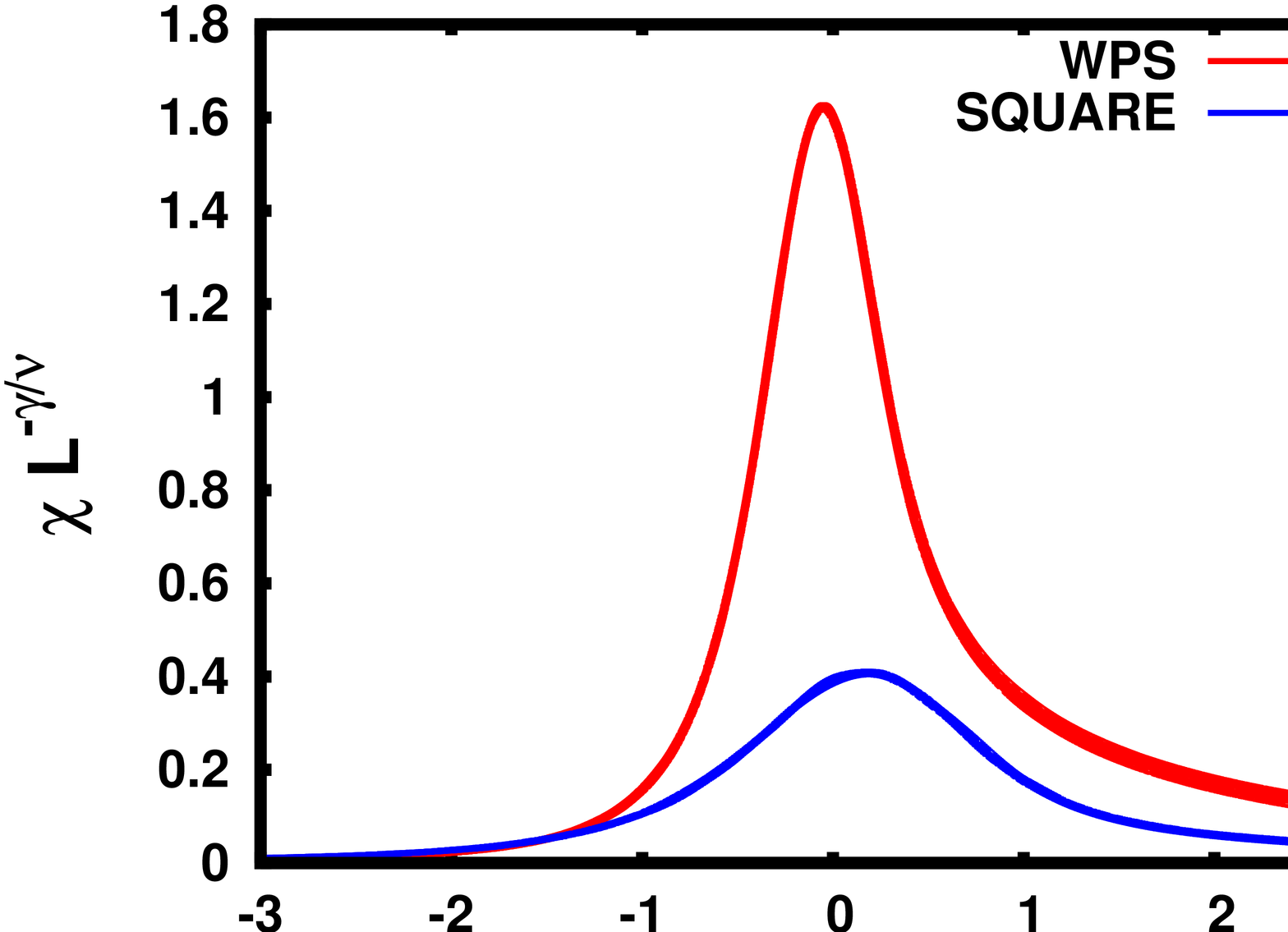}
\label{fig:3d}
}

\caption{Plots of $\chi$ vs $p$ for (a) square and (b) WPS lattices where data represents
convolution of ensemble average of $10,000$ independent realizations. (c) The slopes of the plots of $\log(\chi_h)$ vs $\log(L)$ gives
 $\gamma/\nu=0.635(2)$  and $\gamma/\nu=0.460(1)$ for square and WPS lattices respectively.
In (d) we plot $\chi L^{-\gamma/\nu}$ vs $(p-p_c)L^{1/\nu}$ and find the same data as in (a) and (b) collapse 
into their respective universal curves. 
} 
\label{fig:4abcd}
\end{figure}

Recall that the order parameter $P$ of percolation is defined
as the the ratio of the size of the largest cluster $s_{{\rm max}}$ to the lattice size $N$.
We then keep record of the successive jump size in $P$ i.e., $\Delta P$ within the successive interval
$\Delta p={{1}\over{M}}$ where $M$ is the total number of bonds. The idea of successive jump size in $P$ 
was first studied by Manna in the context of explosive percolation \cite{ref.manna}.
We, however, here consider the ratio ${{\Delta P}\over{\Delta p}}$ and define it as the
susceptibility $\chi(p)$.
In Figs. (\ref{fig:3a}) and (\ref{fig:3b}) we plot $\chi(p)$  as a function 
of $p$ for both types of lattice and find that $\chi(p)$ grows as we increase $p$
and as we approach $p_c$ the growth is quite steep. However, beyond $p_c$ it decreases sharply 
without excluding the size of the spanning cluster. This is in sharp contrast 
to the existing definition of susceptibility as we know that the mean cluster size decreases beyond 
$p_c$ only if the spanning cluster size
is excluded. To find the exponent $\gamma$, we apply the FSS hypothesis.
Following the same procedure done for $C(p)$, we 
find $\gamma/\nu=0.635(2)$ for square lattice and $\gamma/\nu=0.460(1)$ for WPS lattice 
(see Fig. (\ref{fig:3c})). Plotting $\chi L^{-\gamma/\nu}$ versus   $(p-p_c)L^{1/\nu}$ 
we find that all the distinct plots of Figs. (\ref{fig:3a}) and (\ref{fig:3b}) collapse into their own distinct universal curve 
as shown in Fig. (\ref{fig:3d}). It implies that   
\begin{equation}
\chi(p)\sim (p-p_c)^{-\gamma},
\end{equation} 
where $\gamma=0.846(2)$ and $\gamma=0.750(6)$ for square and WPS lattices respectively. It suggests two important
developments. Thus, we find that the susceptibility diverges at the critical point without having to
exclude the spanning cluster. Moreover, the value of the critical exponent $\gamma$ 
is far too less than what we find from the mean cluster size.

\begin{table}[h!]
\centering
    \begin{tabular}{| l | l | l |l|l|}
    \hline
    Lattice & $\alpha$ & $\beta$ & $\gamma$ & $\alpha+2\beta+\gamma$ \\ \hline
    Square & 0.906 & 0.1388 & 0.846 & 2.029 \\ \hline
    WPS & 0.816 & 0.222 & 0.750 & 2.01 \\
    \hline
    \end{tabular}
\caption{The critical exponents and Rushbrooke inequality for random percolation on square and WPS lattices.}
\label{table:1}
\end{table}

To check whether the critical exponents $\alpha, \beta$ and $\gamma$ obey
the Rushbrooke inequality we substitute their values in the Rushbrooke relation and the results are 
shown in table \ref{table:1}.  It clearly suggests that 
Rushbrooke inequality holds in percolation. However, the point to emphasize is that it obeys rather 
more as equality,  within the limits of error,  than as a inequality. Many
experiments and exactly solvable models of thermal CPT too suggest that the Rushbrooke inequality actually holds
as an equality \cite{ref.Gaunt}. Through this work 
we show that the entropy and the order parameter complements each other exactly like in the thermal counterpart. 
For instance, we find that the order parameter $P$, which quantifies the degree of order, is equal to zero 
for $q\geq q_c$ but entropy, which measures the degree of order, is significantly high revealing the high-$q$ 
is the disordered phase exactly like high-$T$ phase in the ferromagnetic transition. On the other hand, at 
$q<q_c$ we find that the entropy $H$ is negligibly small where $P$ increases with 
decreasing $q$ to its maximum value at $q=0$ revealing that low-$q$ is the ordered phase which is
once again like low-$T$ phase in the ferromagnetic transition.
It all implies that percolation is indeed an order-disorder transition.

To summarize, in this article we proposed the equivalent counterpart of entropy and specific heat and 
re-defined the susceptibility 
for percolation model. To measure entropy for percolation we proposed cluster picking probability (CPP)
and shown that the  Shannon entropy for CPP has the same generic feature as the thermal entropy. 
Until now, we could only quantify the extent order of the ordered
phase by measuring $P$ and we could say nothing about the other phase since $P=0$
 there. Now, having known entropy, we can also quantify the other phase and regard
it as the disordered phase as entropy is high and always keep increasing with $1-p$. We have also shown
that specific heat diverges with positive and susceptibility both exhibit power-law near $p_c$ and divergence at $p_c$ without
having to exclude the spanning cluster from their calculations. Using the FSS hypothesis and the idea of data-collapse, 
we numerically obtained the critical exponents $\alpha$ and $\gamma$ for both the lattices. 
Their distinct values for the two lattices
once again confirm that they belong to two different universality classes. Finally, we showed
that the Rushbrooke inequality holds in percolation on 
both the lattices albeit they belong to two distinct universality classes.

\end{document}